\documentclass[12pt]{article}
\newcommand{\beq}{\begin{equation}}
\newcommand{\eeq}{\end{equation}}
\newcommand{\bra}{\begin{array}}
\newcommand{\era}{\end{array}}

\newcommand{\Om}{\Omega}

\input epsf.sty
\usepackage{latexsym}
\author{Arafa H Aly\footnote{
Permanent address: Physics department, Faculty of Sciences,
Beni-Suef University, Egypt, e-mail:arafa16@yahoo.com} and Jamila Douari\footnote{jdouari@gmail.com}\\ \\
\small\it Center for Advanced Mathematical Sciences\\\small\it
American University of Beirut\\ \small\it P.O.Box 11-0236, College
Hall \\\small\it Beirut, Lebanon\rm}
\title{Superconducting Quantum Point contacts and Maxwell Potential}

\begin{document}
\maketitle \vspace*{0.5cm} PACS: 73.23.-b; 73.23.Ad \vskip1cm
Keywords: Quantum Point Contact, Quantization, Andreev
Refelection,Maxwell Potential. \vspace*{1.5cm}
\section*{Abstract}
\hspace{.3in}The quantization of the current in a superconducting
quantum point contact is  reviewed and the critical current is
discussed at different temperatures depending on the carrier
concentration as well by suggesting a constant potential in the
semiconductor and then a Maxwell potential. When the Fermi wave
length is comparable with the constriction width we showed that
the critical current has a step-like variation as a function of
the constriction width and the carrier concentration.

\section{Introduction}
\hspace{.3in}The superconducting quantum point contact (SQPC) is consisting of a split-gate
superconducting-two dimensions electron gas (2DEG)- superconductor junction \cite{sqpc}.
It has attracted the attention of many authors  theoretically and experimentally from the
 early 1970s starting by the studies of the dc Josephson effect in long superconductor-normal
 metal-superconductor junctions \cite{early}. In general, a quantum point contact is a short
  constriction of variable width, comparable to the Fermi wavelength, defined using a split-gate
  technique in a high-mobility 2DEG. Quantum point contacts \cite{qpc} are best known for their
  quantized conductance at an integer multiples of $e^2/h$. Thus, a ballistic theory leads to
   predicting a steplike structure with the conductor having an amplitude $e^2/h$ as a function
   of Fermi energy or width and the current shows a steplike variation as a function of the
   width of the constriction.\\

\begin{figure}
\begin{center}
\mbox{\epsfysize=5cm
   \epsfxsize=8cm
  \epsffile{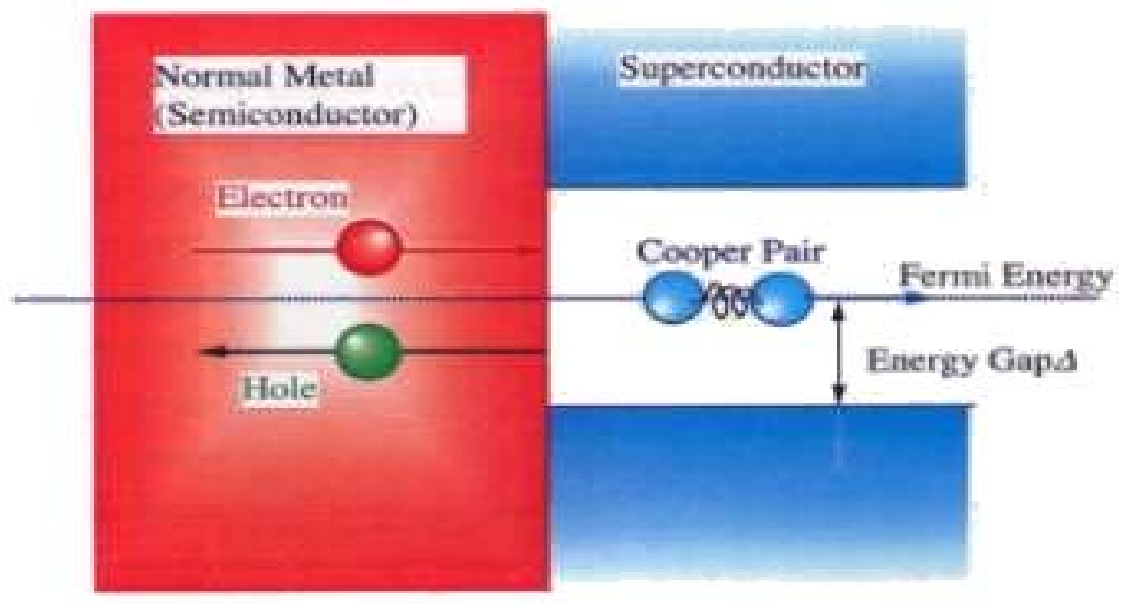}}
\end{center}
\bf{Figure 1: An incident electron from the normal metal is reflected as a hole.
}\rm
\vskip 1cm
\end{figure}
It was found that the Josephson current increases stepwise as a function of the constriction width \cite{j1,j2}, while this current shows oscillations \cite{j1,j4} as a function of the carrier concentration of the 2DEG in the semiconductor layer. This oscillation is due to the interference effects of the quasiparticles that undergo Andreev as well as normal reflection. These results are a characteristic of the transport across a junction with high probability of Andreev reflections.\\

The transmission of quasi-particles through superconductor-normal metal (SN) interfaces requires conversion between dissipative currents and dissipationless supercurrents and is made possible by a two-particle process known as Andreev reflection (AR) (Figure 1) \cite{AR}. An electron injected from the normal metal with energy lower than the superconductor gap is reflected as a phase-matched hole, while a cooper pair is transmitted in the superconductor. Due to its two-particle nature, AR is strongly affected by the transmissivity at the SN interface and much effort has to be devoted to the optimizing of this parameter \cite{AR2}. In the presence of scattering centers in the normal region, the phase relationship between incoming and retroreflected particles can give rise to marked coherent-transport phenomena such as reflectionless tunneling \cite{AR3}.\\

In this paper, we consider non-zero temperature for the dc
Josephson effect of SQPC's by suggesting a constant potential in
the semiconductor in subsection 2.1, then a Maxwell potential
which could be gotten from a contact of the system with a photon's
source in subsection 2.2. We deal with the influence of the
presence of photons on the SQPC. Theoretically, we make use the
pure Maxwell theory which assures the anyonic properties of 2DEG
and avoids the appearance of topological mass that is responsible
for the screening characteristics displayed by pure electric
charges in the electrodynamics controlled by the
Maxwell-Chern-Simons theory \cite{mcst}. In two dimensions it is
allowed the possibility of particles with any statistics, where
the physical excitations obeying it are called anyons \cite{an}.
Thus, a concrete way to realize non-trivial statistics is by
attaching a magnetic flux to electrically charged particles
forming a composite system. In this context, we discuss the
influence of the extra potential that we call Maxwell potential on
the critical current at different temperatures for our pruposed
system and we examine the dependence of the carrier concentration
on the current.

\section{Methodology}
\subsection{Josephson Current Through a Quantum Point Contact}
\hspace{.3in}A quantum semiconductor device comprises mainly of a
channel region
 formed with a two-dimensional carrier gas. A Schottky electrode structure
  is provided on the channel region for creating a depletion layer in the channel
  region to be extended in a lateral direction such that the two-dimensional carrier
  gas is divided into a first and a second region. A quantum point contact
   formed in the depletion layer to connect the first and second region of the
   two-dimensional carrier gas in a longitudinal direction; an emitter electrode
  is provided on the channel region in correspondence to the first region of the
     two-dimensional carrier gas; one or more collector electrodes are provided on
     the channel region in correspondence to the second region of the two-dimensional
     carrier gas, and another Schottky electrode structure is provided in correspondence
     to the first region for creating a depletion region therein, such that a path of
     the carriers entering into the quantum point contact is controlled asymmetrical
     with respect to a hypothetical longitudinal axis, that passes through the quantum
      point contact in the longitudinal direction.\\

Due to the generation of depletion layer \cite{j4,j3}, the width of
the constriction $W_n$ is reduced to the following (see figure 2)
when an applied voltage is biased, \beq W_{n'}=W_{n}-2\rho \eeq The
depletion layer $\rho$ depends on the bias voltage $V_0$, the
carrier concentrations $N_B$, the temperature $T$ and the
permittivity of the material $\epsilon$\cite{j6}, and it is given
by
\beq \rho=\sqrt{\frac{2\epsilon}{eN_B}(V_0 -\frac{2k_B T}{e})}.
\eeq
\begin{figure}
\begin{center}
\mbox{\epsfysize=7cm
   \epsfxsize=7cm
  \epsffile{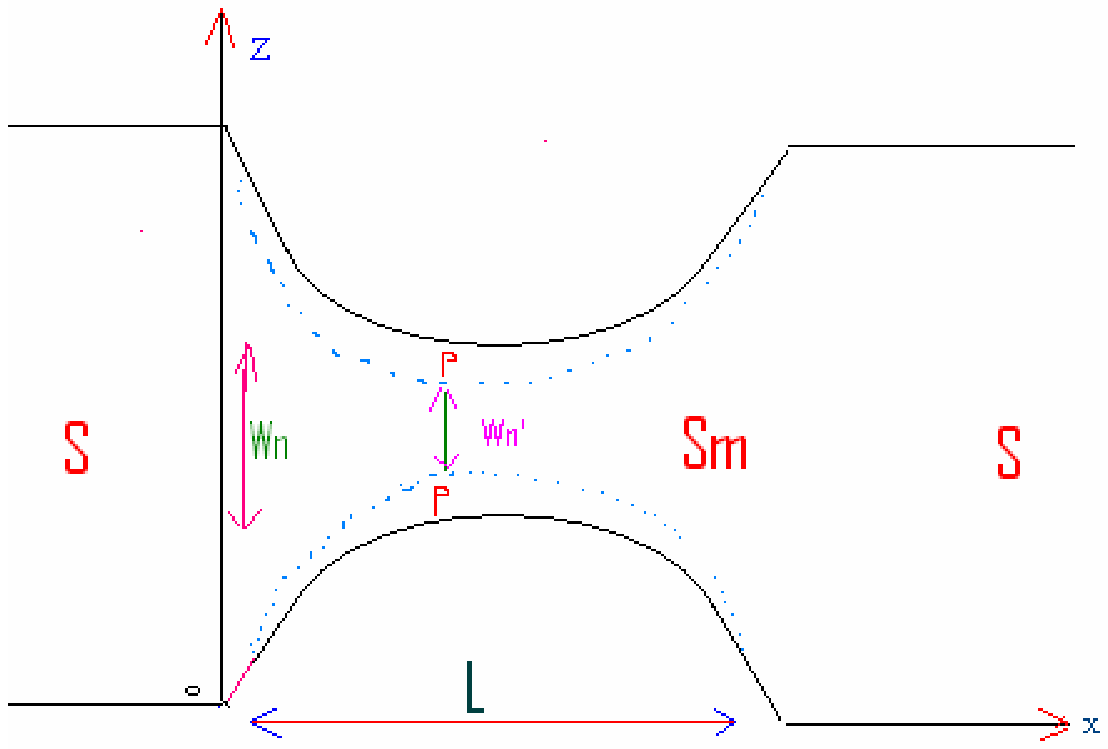}}
\end{center}
\bf{Figure 2: Superconducting Quantum Point contact.
}\rm
\vskip 1cm
\end{figure}
The device geometry assumed for our calculation is shown in Fig.2.
Our system is described by the Bogoliubov-de-Gennes (BdG) equation
\cite{j6}, which is given by \beq
\pmatrix{-H(x,y)+U(x,y)&\Delta(x,y)\cr\Delta^*(x,y)&H(x,y)-U(x,y)\cr}\pmatrix{u(x,y)\cr
v(x,y)\cr}=E\pmatrix{u(x,y)\cr v(x,y)\cr}, \eeq Solutions to this
equations are electron-like and hole-like quasiparticles (QP) wave
functions \cite{j2,j4}; $u(x,y)$ and  $v(x,y)$ represent the
eigenfunctions for the electron and hole quasiparticles .

We consuder a simple model of a Josephson junction that shows the
essential features of the Josephson Effect. It is a
one-dimensional model, where the left and right superconductors
have the pair potentials of the same magnitude but with different
phases \beq \Delta(x)=\left\{\bra{ll}
\Delta_0 e^{i\theta_L}\phantom{~~~~}x<0\\
\Delta_0 e^{i\theta_R}\phantom{~~~~}x>0
\era\right.
\eeq

We assume the normal region to be thin and we consider an extreme
case where its width is infinitesimal. In general, some scattering
process is expected to be present either inside the normal region
or at the super-normal interfaces; Introducing a scattering
potential. \beq U(x)=V_b+\mu_L,_R+U_0 \eeq with $V_b$ representing
the height of the Schottky barrier at the S-Sm interface, $\mu$
the chemical potential, and $U_0$ the potential energy of the
interface.

The basic wave function of the quasiparticles in the jth channel
can be written as \beq
\Psi_j(x,y)=\sqrt\frac{2}{W_{n'}}\phi_j(x)\sin(j\pi[\frac{y}{W_{n'}}+\frac{1}{2}]),
\eeq where a two-component wave function $\phi_j(x)$ obeys a
modified BdG equation \beq
\pmatrix{\frac{-\hbar^2}{2m^*}(\frac{\partial^2}{\partial
x^2}-(\frac{j\pi}{W_{n'}})^2)+V_b+\mu_L,_R+U_0&\Delta(x)\cr
\Delta^*(x)&\frac{\hbar^2}{2m^*}(\frac{\partial^2}{\partial
x^2}-(\frac{j\pi}{W_{n'}})^2)-V_b-\mu_L,_R-U_0}\phi_j (x)=E\phi_j
(x), \eeq

The solution of eqn.(7) is
\beq
\phi_j(x)=\left\{\bra{lll}
e^{iP^+_j x}\pmatrix{u_0\cr v_0 \cr} +a_{1j}e^{iP^-_j x}\pmatrix{u_0\cr v_0 \cr}+b_{1j}e^{-iP^+_j x}\pmatrix{u_0\cr v_0 \cr} \phantom{~~~~~~~~~~~~}x<0\\
c_{1j}e^{iP^+_j x}\pmatrix{u_0 e^{i\phi}\cr v_0 e^{-i\phi}\cr}
+d_{1j}e^{iP^-_j x}\pmatrix{u_0 e^{i\phi/2}\cr v_0
e^{-i\phi/2}\cr}\phantom{~~~~~~~~~~~~~~~~~~~~~~~}x>L \era\right.
\eeq where $P^{\pm}_j =\sqrt{\frac{2m^*}{\hbar^2}(V_b+\mu_L,_R+U_0
\pm \Omega -q_j^2)}$,
 $q_j=\frac{j\pi}{W_s}$, $\Omega=\sqrt{E^2 -\Delta^2_0}$ in which $W_s$ is the width of
 the superconducting  electrodes and $k^{\pm}_j =\sqrt{\frac{2m^*}{\hbar^2}(V_b+\mu_L,_R+U_0 \pm E)-(\frac{j\pi}{W_{n'}})^2}$ with $u_0=\sqrt{\frac{1}{2}(1+\frac{\Om}{E})}$ and $v_0=\sqrt{\frac{1}{2}(1-\frac{\Om}{E})}$.\\
The coefficients $a_{1j}$,$b_{1j}$,$c_{1j}$and $d_{1j}$ are
functions of the Energy $E$ and the phase difference $\phi$ across
two superconductors, they are determined by the matching
conditions at the interfaces. We solve the BdG equation in the WKB
approximation\cite{j7} to obtain the amplitude $a_{1j}$, which is
given by \beq
a_{1j}(\phi,E)=\frac{\Delta_0[e^{i\theta_j}-e^{-\phi}]}{(E+\Om)e^{-\phi}-(E-\Om)e^{i\theta_j}}
\eeq
Where $\theta_j=L(k^+_j -k^-_j)$.\\

The dc Josephson current, $I$, due to Andreev reflections can be
calculated as \beq
I=\frac{e\Delta_0}{\hbar\beta}\sum\limits_{w_n}\frac{1}{\Om_n}\sum\limits_{j=1}^{N}[a_{1j}(\phi,iw_n)-a_{1j}(-\phi,iw_n)],
\eeq where $w_n$ is the Matsubara frequency,given by
$w_n=\pi(2n+1)/\beta$ with $\beta=(k_B T)^{-1}$ and $T$ the
absolute temperature, $k_B$  the Boltzman constant,and finally
$\Om_n=\sqrt{w_n+\Delta^2}$. Now substituting equation (9) into
equation(10), we get for the current $I$ \beq
I=\frac{2e\Delta^2_0}{\hbar\beta}\sum\limits_{w_n}\sum\limits_{j=1}^{N}
\frac{\sinh \phi}{(2w_n^2 +\Delta^2_0)\cosh\tilde{\theta}_j
+2w_n\sqrt{w_n^2 +\Delta^2_0}\sinh\tilde{\theta}_j +\Delta^2_0
\cos\phi}, \eeq
where $\tilde{\theta}_j=-i\theta_j (E\longrightarrow iw_n)$.\\

\begin{figure}
\begin{center}
\mbox{\epsfysize=5cm
   \epsfxsize=5cm
  \epsffile{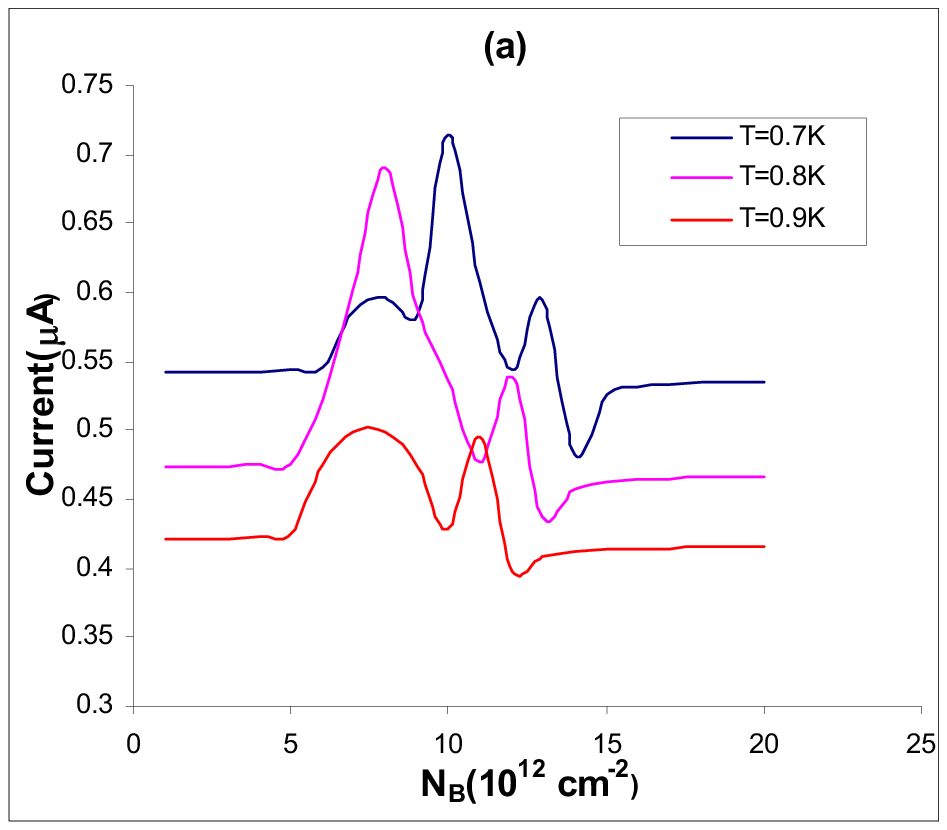}}
  \mbox{\epsfysize=5cm
   \epsfxsize=5cm
  \epsffile{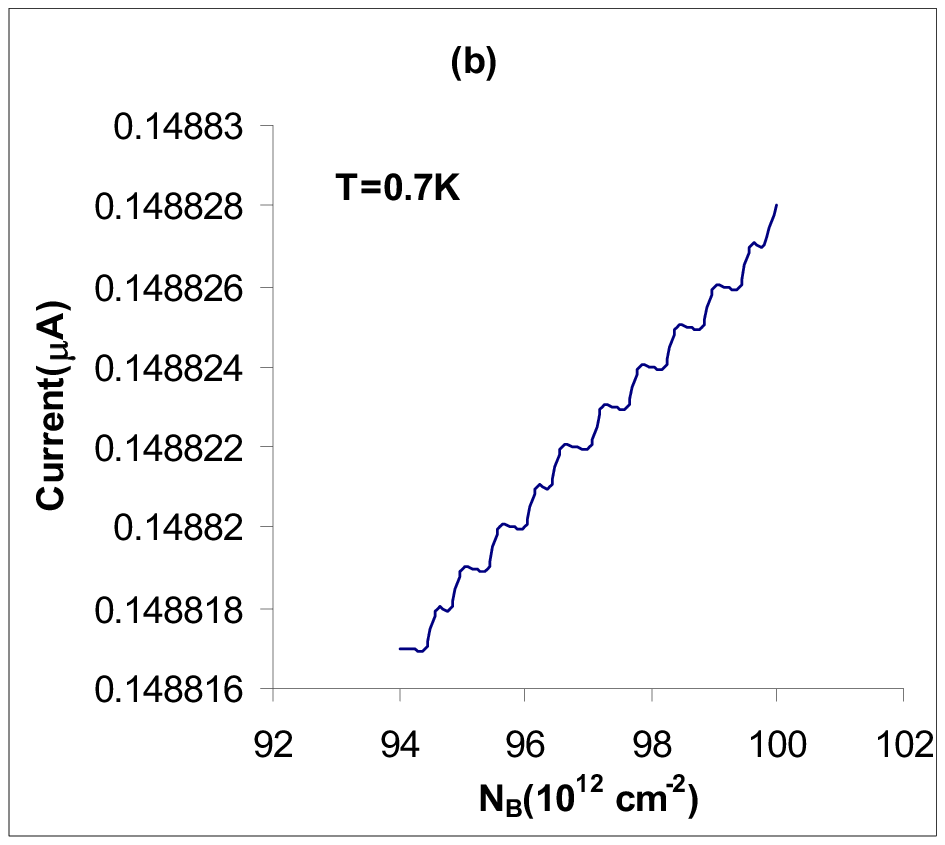}}
\end{center}
\vskip 1cm \bf{Figure 3: The variation of the current with the
doping concentration at Wn=3nm and $\phi =2\pi$. (a) Small scale.
(b) Large scale. }\rm
\end{figure}

\begin{figure}
\begin{center}
\mbox{\epsfysize=5cm
   \epsfxsize=5cm
  \epsffile{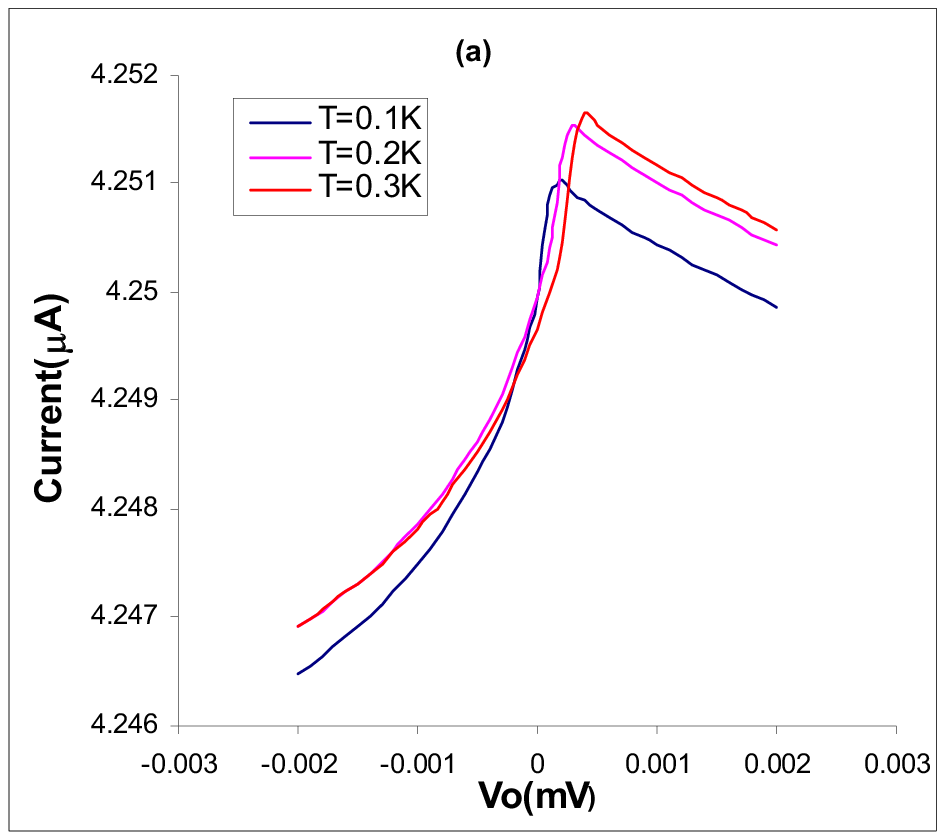}}
  \mbox{\epsfysize=5cm
   \epsfxsize=5cm
  \epsffile{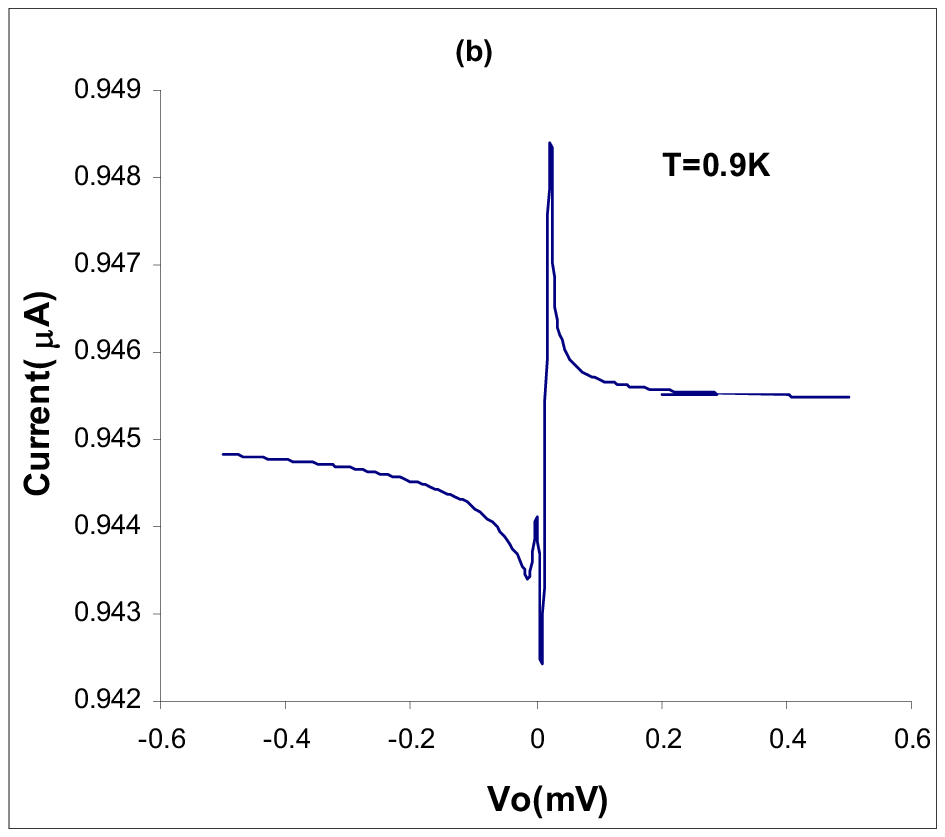}}
\end{center}
\bf{Figure 4: I-V characteristics at Wn=13nm and $\phi =2\pi$. (a)
Small scale. (b) Large scale. }\rm \vskip 1cm
\end{figure}

\begin{figure}
\begin{center}
\mbox{\epsfysize=5cm
   \epsfxsize=7cm
  \epsffile{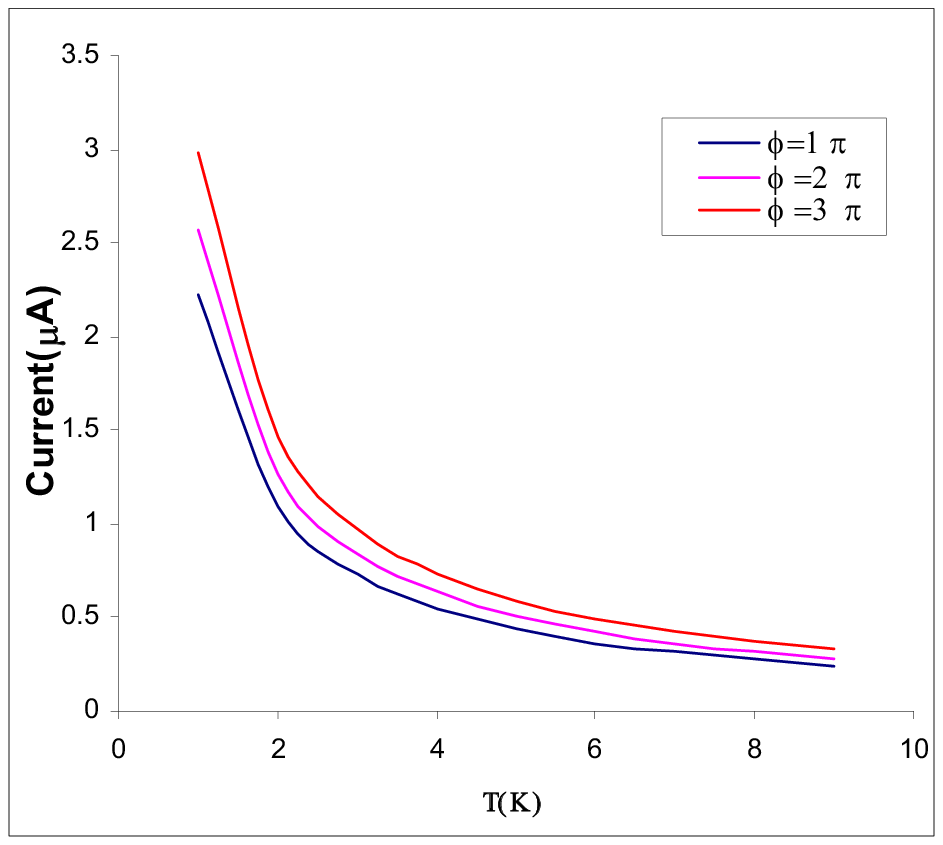}}
\end{center}
\bf{Figure 5: Dependence of the current on the temperature at different
phases and constriction =3 nm.
}\rm

\end{figure}
\vskip 10cm
\subsection{Josephson current and Maxwell Potential}
\hspace{.3in}In this section we would like to discuss the influence of the presence of photons on the superconducting quantum point contact in which the 2DEG is settled as weak region linking two superconductings. We make use the pure Maxwell theory which assures the anyonic properties of 2DEGA concrete way to realize non-trivial statistics is by attaching a magnetic flux to electrically charged particles forming a composite system. Theoretically the system is described by the Lagrangian
\beq
\emph{La}=-\frac{1}{4}F_{\mu\nu}F^{\mu\nu}+J^\mu A_\mu \eeq The Lagrangian describes Maxwell theory that couples to the current. In this theory, the gauge fields are dynamic and the canonical moments are $\pi^\mu =F^{\mu 0}$ which results in the usual primary constraint $\pi^0 =0$ and $\pi^i =F^{i 0}$. Thus for a composite located at the origin, the solution of the Maxwell equations read $B(x)=\xi\delta^2(x)$ and $E^i(x) =-\frac{e}{2\pi}\frac{x^i}{r^2}$ with $r=|x|$ and $\xi$ is the dipole's moment \cite{dm}.\\

The composite system in this theory interacts with purely maxwellian photons and displays a potential with the confining nature \beq U(r)=\frac{e^2}{\pi}\ln(\eta r)\eeq with $\eta$ is a massive cutoff \cite{mcst}, for simplicity we take $\eta=1$. As a remark, the potential agrees with the behavior of the Maxwell-Chern-Simons theory in the limit of short separation.\\

Now, considering a superconducting quantum point contact which is
compatible with our purely Maxwell theory in 2-dimension, the
potential given in the previous section will be changed to the one
depending on the distance separating the electrons in 2DEG region
$0<x<L$ and $\forall z$ (fig.2) leading to an important change in
the Josephson current of the SQPC as will be shown below. We
account for the the different distances between the chemical
potential and conduction band edge in the S and Sm regions,
respectively, by introducing a potential step of height $U_0$. The
potential can be written as follows $U(r)=U_1(x)+U_2(z)$. Since
the Nb are modeled as half-infinite slabs of thickness $W$
occupying the superconducting regions. Then, \beq
U_2(z)=\left\{\bra{ll}
0,&0<z<W\\
\infty,&z<0, z>W
\era\right.
\eeq
this for $x<0$ and $x>L$. If $0<x<L$ and for a typical value $W=100nm$ the potential is
\beq
U_2(z)=\left\{\bra{ll}
\infty,&z<0\\
eF_s z&,0<z
\era\right.
\eeq
This approximates the potential of an inversion layer, the surface electric field being given by $F_s$. Now we model the potential in the direction of $x$ as
\beq
U_1(x)=U_0 +V_b +\frac{e^2}{\pi}\ln(x),
\eeq
the first and second terms are the potentials introduced in the previous section.\\

By replacing the new potential $U(x)$ by (16) in the equation (7) the Josephson current will be affected as we will see below.\\

We solve the equation (7) with the new potential (16) is
considered and we get that the function $\phi_j$ doesn't chage for
$x<L$ and $x>L$. If $0<x<L$ the egenfunction becomes \beq \phi_j
(x)=\Big( c_1 A_i(X)+c_2B_i(X)\Big)\pmatrix{0\cr1\cr}+\Big( c_3
A_i(Y)+c_4B_i(Y)\Big)\pmatrix{1\cr0\cr}, \eeq with $$X=\frac{\pi
j^2}{W_n^2}(\frac{\pi\hbar^2}{2e^2
m})^{\frac{2}{3}}-(x-1)(\frac{\pi\hbar^2}{2e^2
m})^{\frac{-1}{3}}$$ $$Y=\frac{\pi
j^2}{W_n^2}(\frac{\pi\hbar^2}{2e^2
m})^{\frac{2}{3}}+(x-1)(\frac{\pi\hbar^2}{2e^2
m})^{\frac{-1}{3}}$$ and $A_i(X)$, $B_i(X)$ are Bessel functions
of first and second kinds respectively. By matching the conditions
at $x=0$ and $x=L$ within the Andreev and WKB approximation using
(8) and (17) we get $b_{1j}=c_2=c_4=d_{1j}=0$ and \beq
a_{1j}(\phi)=\frac{u_0}{v_0}\frac{P-1}{1-\frac{u^2_0}{v^2_0}P}\eeq
where
$$P=\frac{A_i(Z_1^+) A_i(Z_2^-)}{A_i(Z_1^-) A_i(Z_2^+)}e^{i\phi}$$
and $Z_1^\pm =a\pm b$ and $Z_2^\pm =a\pm b(\frac{L^2}{2}-1)$ in which $a=j^2 (\frac{\hbar^2}{2e^2 m})^{2/3}\frac{\pi^{5/3}}{W_n^2}$ and $b=2^{2/3}(\frac{\pi\hbar^2}{e^4 m^2})^{-1/3}$. To determine the exact expresion for $P(X)$ we see the following approximations. If $X>0$ (which is consistent with our values) the Bessel function of first kind is reduced to $$A_i(X)=\frac{\sqrt{\frac{x}{3}}}{\pi}K_{1/3}(\frac{2}{3}X^{\frac{2}{3}})$$
and again if $$\frac{2}{3}X^{\frac{2}{3}}=Z>\frac{1}{3}$$ the modified Bessel function $K$ is reduced to $$K_{1/3}(Z)=\sqrt{\frac{\pi}{2Z}}e^{-Z}.$$ Consequently, The function $P$ becomes
$$P=\frac{Z_1^- Z_2^+}{Z_1^+ Z_2^-}e^{i\phi -\frac{2}{3}\Big( (Z_1^+)^{\frac{3}{2}}+(Z_2^-)^{\frac{3}{2}}-(Z_1^-)^{\frac{3}{2}}-(Z_2^+)^{\frac{3}{2}}\Big)}.$$ Now following the definition of Josephson current given by (10) and the expression we obtain for $a_{1j}$ we get
\beq
I=\frac{e\Delta_0}{\hbar\beta}\sum\limits_{w_n}\frac{1}{\Om_n}\sum\limits_{j=1}^{N}\frac{u_0}{v_0}\Big( \frac{P(\phi)-1}{1-\frac{u^2_0}{v^2_0}P(\phi)}-\frac{P(-\phi)-1}{1-\frac{u^2_0}{v^2_0}P(-\phi)}\Big).
\eeq
In our numerical calculations, we dealed with the obtained critical current in terms of different variables at non-zero temperatures as we see on the curves below;

\begin{figure}
\begin{center}
\mbox{\epsfysize=5cm
   \epsfxsize=5cm
  \epsffile{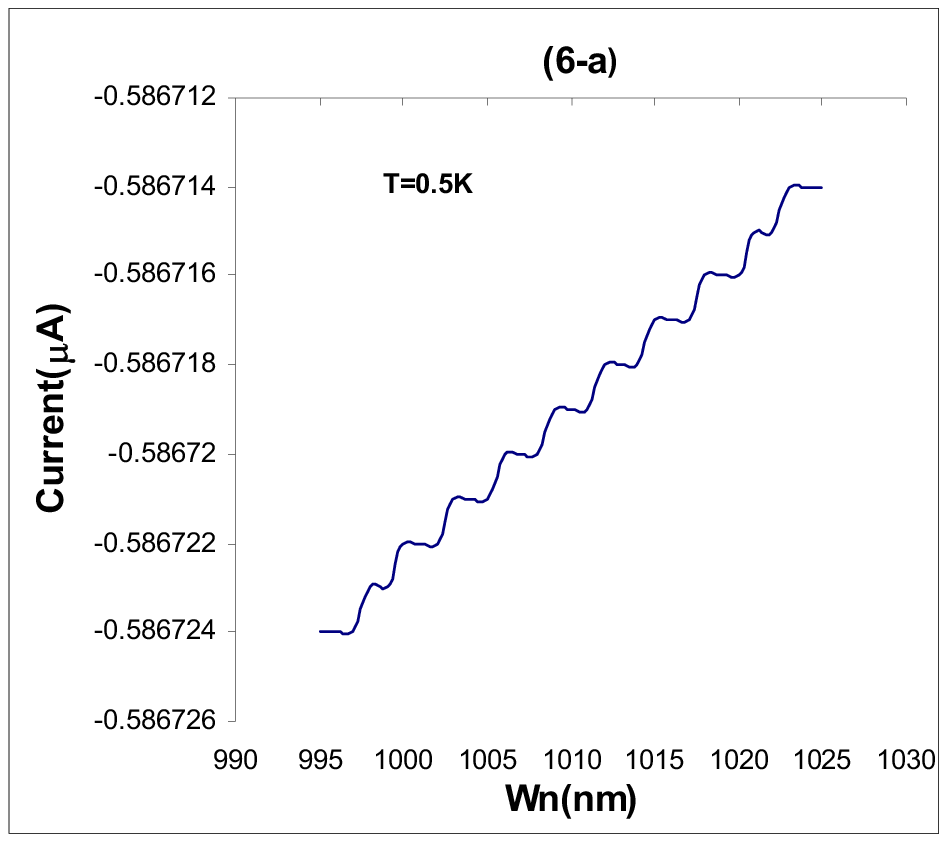}}
\mbox{\epsfysize=5cm
   \epsfxsize=5cm
  \epsffile{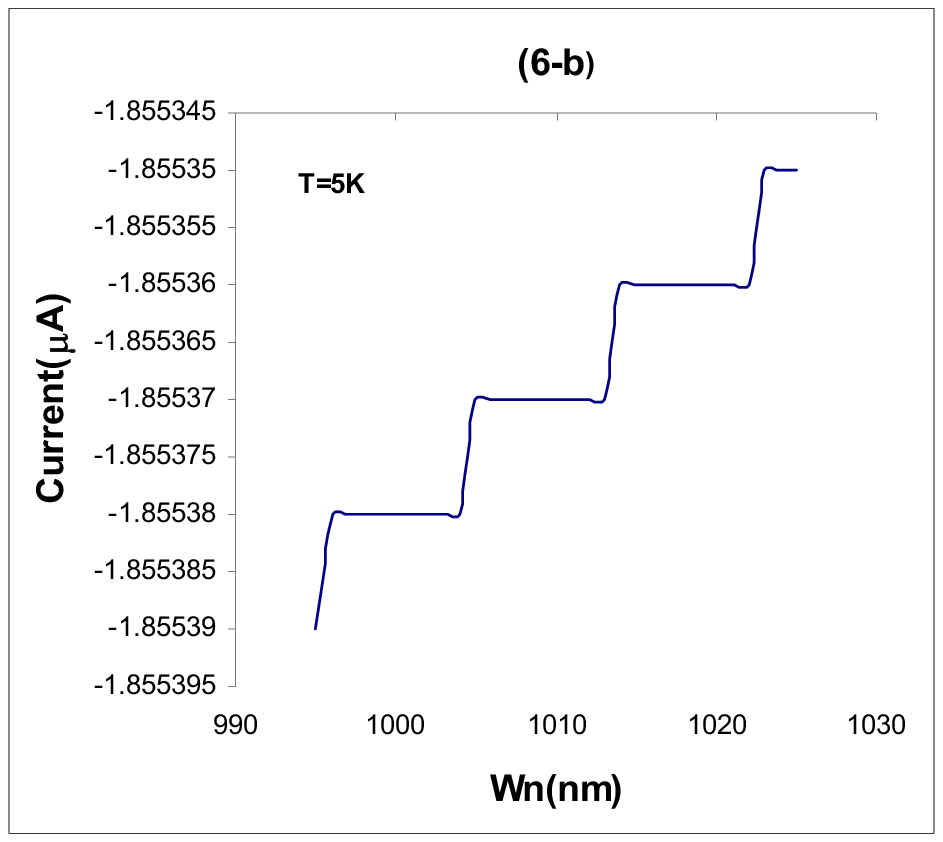}}
\mbox{\epsfysize=5cm
   \epsfxsize=5cm
  \epsffile{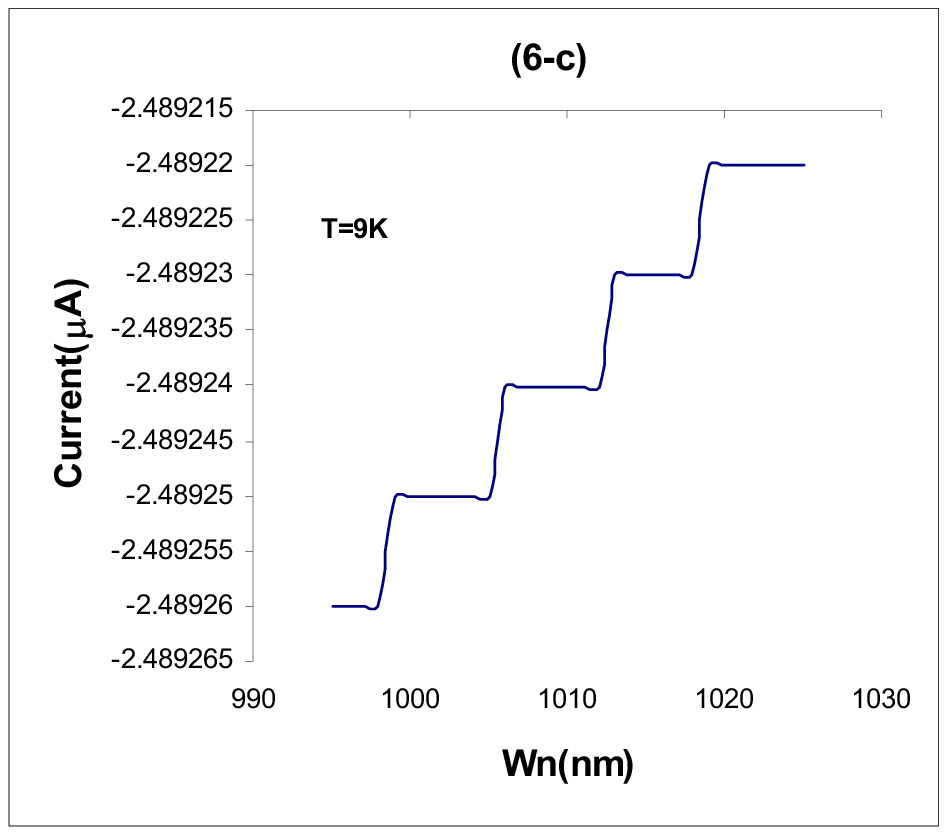}}
\end{center}
\bf{Figure 6: The current variation in terms of large constriction width with $\phi =4.5\pi$.
}\rm
\vskip 1cm
\end{figure}

\begin{figure}
\begin{center}
\mbox{\epsfysize=7cm
   \epsfxsize=7cm
  \epsffile{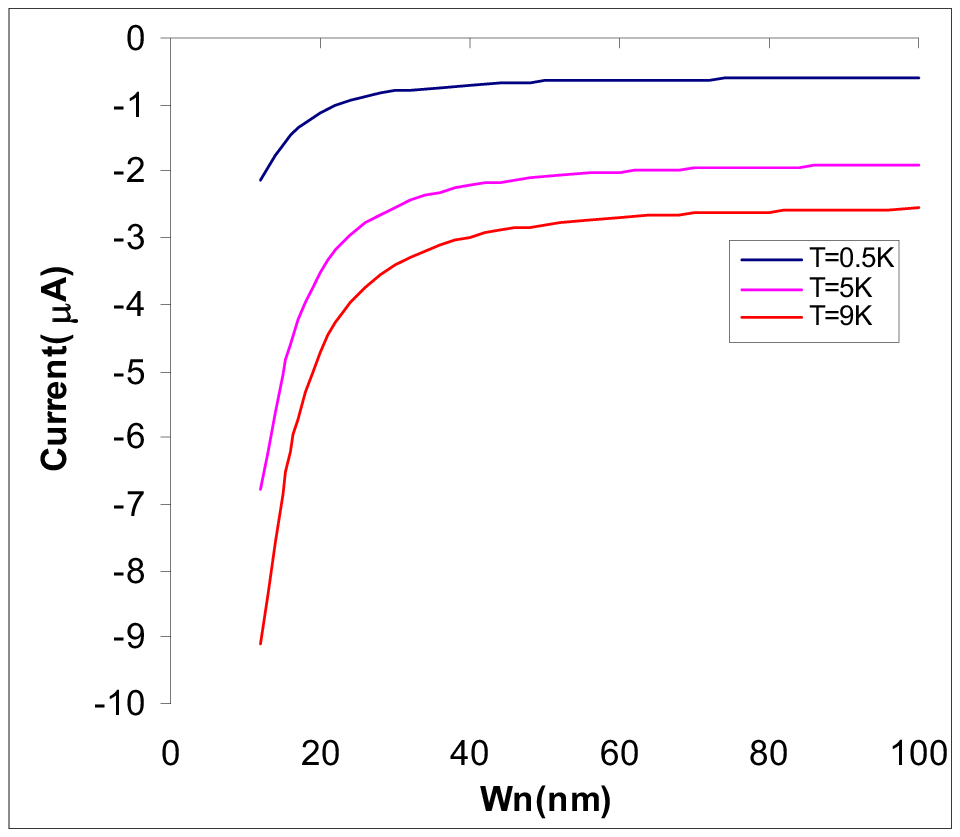}}
\end{center}
\bf{Figure 7: The current variation in terms of small constriction width with $\phi =4.5\pi$.
}\rm
\end{figure}

\begin{figure}
\begin{center}
\mbox{\epsfysize=5cm
   \epsfxsize=5cm
  \epsffile{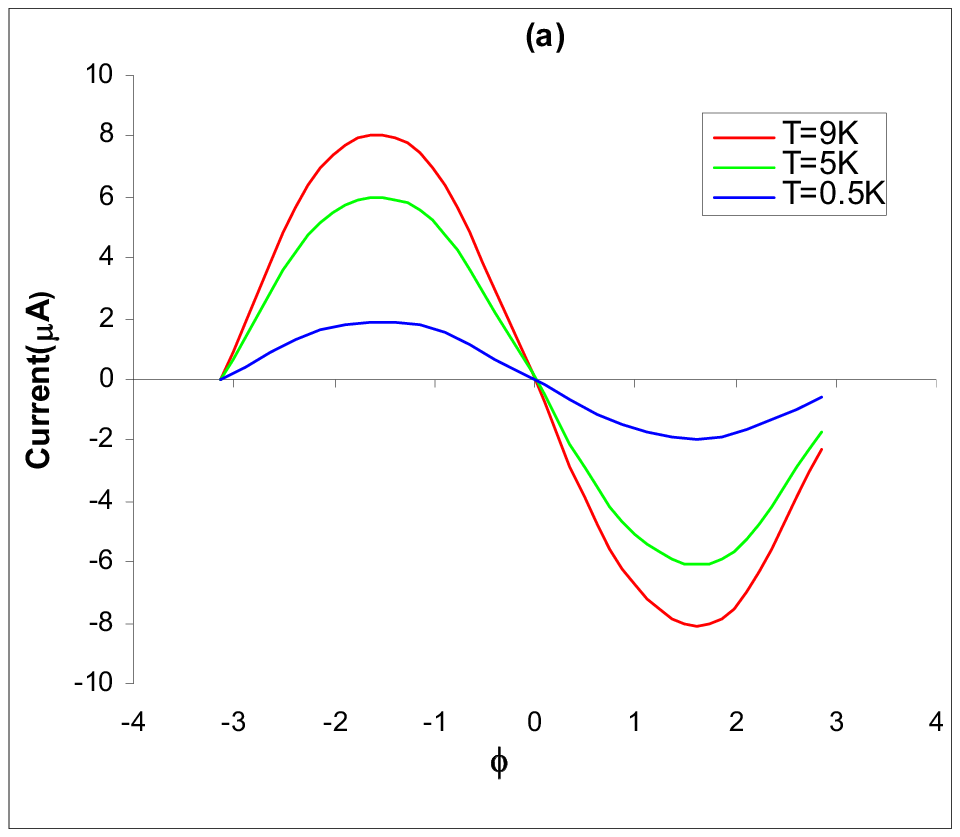}}
\mbox{\epsfysize=5cm
   \epsfxsize=5cm
  \epsffile{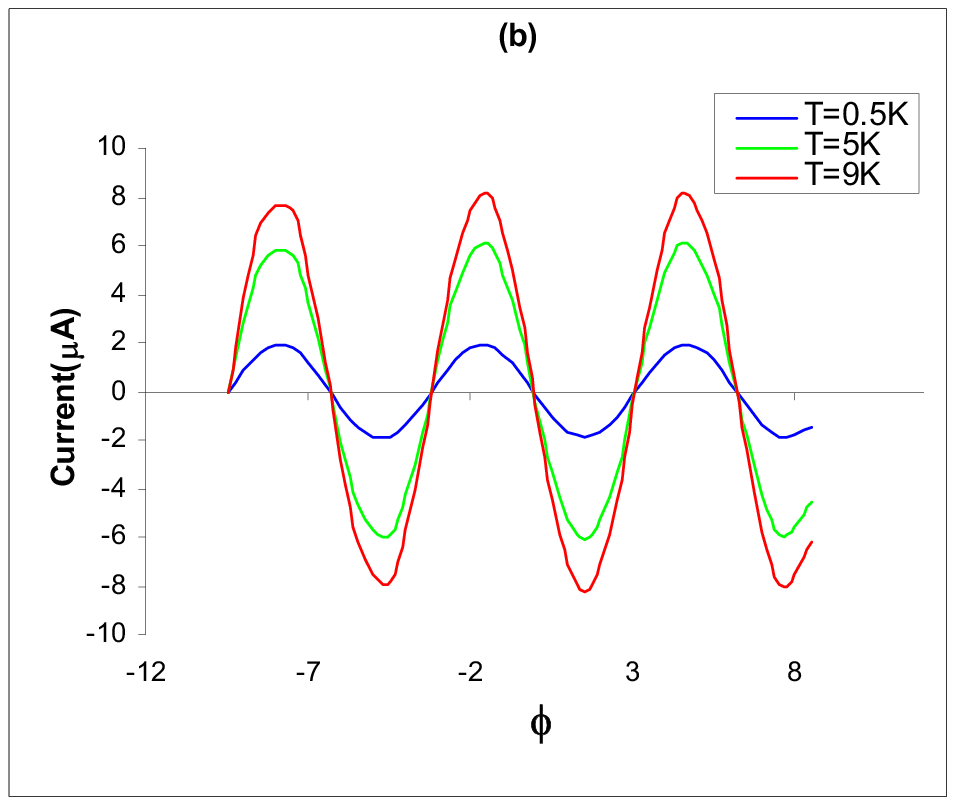}}
\end{center}
\bf{Figure 8: Current-Phase relationship. (a) $\phi$ from $-\pi$ to $\pi$. (b) The range of $\phi$ from $-3\pi$ up to $3\pi$.
}\rm
\end{figure}

\section{Discussion and Conclusion}
\hspace{.3in}We studied the quantization of the supercurrent of a SQPC in a S-2DEG-S Josephson junction with a split gate. The supercurrent  values change stepwise as a function of the carrier concentration and the constriction width. We observe the onset of the first transport mode contributing both to the supercurrent of the SQPC. Furthermore the steps in the supercurrent appear at the same gate voltage values. This shows that each transport mode in the SQPC contributes to the supercurrent.\\

Thus, in order to show the reliability of the present theoretical treatment for the present model superconducting quantum point contact (SQPC) in a superconductor-two dimensional electron gas-superconductor(S-2DEG-S), we have performed a numerical calculation. In real system, however, there always exists Schottky barriers at the S-Sm inetrfaces which reduce the density of the Cooper pairs in the semiconductor \cite{j1}. The electron transport through the junction is treated as a stochastic process, so that the tunneled electron energy as a random number. The Schottky barrier height, $V_b$, is determined by using the Monte-Carlo simulation technique and its value was found to be $\sim 0.49$eV for the case Nb-GaAs Nb-GaAs based heterostructure interface. This value of, $V_b$, was found in agreement with those found experimentally and theoretically \cite{b}.\\

Figures (3a,3b) Show the variation of the current $I$ with the carrier concentration $N_B$. From these figures, it is shown that the current quantizes but when the carrier concentration increases the variation of the current changes and it makes many peaks with different dips. The shape of these dips depends on the value of the temperature and the carrier concentration (see fig.3b), and at the large scale of $N_B$ the special property of the quantum point contact is that upon widening the opening the current does not increase gradually but stepwise when the $N_B$ is increased which can be seen in agreement with the result given in \cite{j3,j5}.\\

In Figure (4a) we found  the bias voltage $V_o$ at different temperatures. As shown from this figure that the dip height at zero bias voltage increases as the temperature increases, and at the large scale (see fig 4b) of bias voltage we found subgab about zero bias which is agree with the result found in the reference \cite{ref,ref2}. Figure (5) shows the decrease of the current $I$ as the temperature T increases at different phases $\pi$, 2$\pi$ and 3$\pi$. This result shows a qualitative agreement with those published in the literature \cite{tem}. This variation shows that Josephson effect is optimal observed at very low $T$, also when the value of $\phi$ increases the value of the current increases too.\\

As shown in ref. \cite{j1,j2}, the current quantization of the
superconducting quantum point contact has been proved and these
results verify the existence of the interference effects of the
quasiparticles that undergo Andreev reflections. In two dimensional
systems, the numerical calculations manipulating the carrier density
via the field effect, verify experimentally this resonance
\cite{j3}.
 Because of the exponential decay of a coherence length, the short junctions and the low temperature
  favor large amplitudes (Figs 4a,4b).
   It is shown that the critical current is quantized and increased stepwise as a function of the width of the semiconductor layer and the doping concentration up to certain values of both.\\

Now, at the presence of extra potential that we called Maxwellian one, in the figures (6,7), we again see that the current $I$ exhibits a peak as a function of the width of constriction $W_n$. Also the special property of the quantum point contact didn't change; upon widening the opening the current does not increase gradually but stepwise when the width is increased (see fig 7), when a steps do occur, the step high depends sensitively on the parameters of the junction and at the large scale of $W_n$ with the low value of $T=0.5K$ the stepwise is seen in figures 6a, 6b and 6c but we remark that when the temperature is increased to 5K and 9K we can see the current is increased with clear stepwise, also in this case our result is in good agreement with what was found in \cite{ref3}. A periodic variation of $I$ with $\phi$ is shown in Figures (8a,8b). This result was observed by another authors \cite{c,phas} previously which shows the coherent property of our present system and it is in a clean limit. Also, the relation $I(\phi)$ (Fig. 8a,8b) for a width 13nm and different temperatures (T=0.5K ,5K ,9K) shows a behavior which is similar to the behavior of the current found in the recent work \cite{c}. This result confirms the reliability of our treatment for the model concerning SQPC. In other words, the quasi-particle reflection from the edgs of the Andreev gap causes mesoscopic phenomena manifested in oscillating features on $I(N_B)$ and $I(W_n)$ and $I(\phi)$.\\

Apart from studying fundamentals of charge transport in mesoscopic conductors, quantum point contacts can be used as extremely sensible charge detectors. Since the conductance through the contact strongly depends on the size of the constriction, any potential fluctuation (for instance, created by other electrons) in the vicinity will influence the current through the QPC. It is possible to detect single electrons with such a scheme. In view of quantum computation in solid-state systems, QPC's may be used as readout devices for the state of a qubit.\\

We have studied the Jospheson effect in SQPC's in the ballistic regime using simple two-dimensional modes. We have found that in some cases the critical current shows a characteristic feature due to the discreteness of energy levels around the constriction. This will be a common feature of the phase coherent ballistic conduction of supercurrent through a constriction in superconducting quantum devices.

\section*{Acknowledgements}
\hspace{.3in}The authors are Arab regional Fellow at CAMS supported by a Grant from the Arab Fund for Economic and Social Development.

\end{document}